**Title**

A data-Oriented based Self-Calibration And Robust chemical-shift encoding by using clusterization (OSCAR): Theory, Optimization and Clinical Validation in Neuromuscular disorders

**Authors**

G. Siracusano[1,*], A. La Corte[1], M. Gaeta[2] and G. Finocchio[3]

[1] Department of Electric, Electronic and Computer Engineering, University of Catania, Viale Andrea Doria 6, 95125 Catania, Italy

[2] Department of Biomedical sciences, Dental and of Morphological and Functional images, University of Messina, Via Consolare Valeria 1, 98125 Messina, Italy

[3] Department of Mathematical and Computer Sciences, Physical Sciences and Earth Sciences, University of Messina, Viale F. D'alcontres 31, 98166 Messina, Italy

* Corresponding Author: giuliosiracusano@gmail.com

**Abstract**

Multi-echo Chemical Shift–Encoded (CSE) methods for Fat-Water quantification are growing in clinical use due to their ability to estimate and correct some confounding effects. State of the art CSE water/fat separation approaches rely on a multi-peak fat spectrum with peak frequencies and relative amplitudes kept constant over the entire MRI dataset. However, the latter approximation introduces a systematic error in fat percentage quantification in patients where the differences in lipid chemical composition are significant (such as for neuromuscular disorders) because of the spatial dependence of the peak amplitudes . The present work aims to overcome this limitation by taking advantage of an unsupervised clusterization-based approach offering a reliable criterion to carry out a data-driven segmentation of the input MRI dataset into multiple regions. The idea is to apply the clusterization for partitioning the multi-echo MRI dataset into a finite number of clusters whose internal voxels exhibit similar distance metrics. For each cluster, the estimation of the fat spectral properties are evaluated with a self-calibration technique and finally the fat/water percentages are computed via a non-linear fitting. The method is tested in ad-hoc and public datasets. The overall performance and results in terms of fitting accuracy, robustness and reproducibility are compared with other state-of-the-art CSE algorithms. This approach provides a more accurate and reproducible identification of chemical species, hence fat/water separation, when compared with other calibrated and non-calibrated approaches.

**Keywords:** Fat-Water Separation, Chemical Shift, Clustering, Gap Statistics, Multi-Echo, Multi-Peak Fat Spectrum, Skeletal muscle, Neuromuscular disorder;

## 1. Introduction

The applications of quantitative chemical shift-encoded (CSE) methods for fat/water separation are experiencing a growing interest due to the need of a robust and accurate lipid quantification in different body parts [1–3]. In the case of skeletal muscle, the estimation of fatty infiltration is important in analyzing the progression of neuromuscular disorders (NMDs) (i.e. Duchenne [4–7], myopathies [8]) assessing risk factors and monitoring therapy for metabolic abnormalities (like obesity and diabetes) [9], and in grading muscle degeneration after injuries [10]. The accurate evaluation of proton density fat fraction (PDFF) [11–14] requires the consideration of multiple confounding factors including: the field map variations [13–16], the complexity of fat spectrum [14,17], the effect of $T_2^*$ decay [18,19], the $T_1$-weightening [20,21], the noise-induced bias [22], the eddy currents [16,23], the susceptibility [24] and also temperature effects [25].

Among them, the careful modeling of the fat spectrum complexity has been primarily accomplished by using a multi-peak spectrum model with three main approaches for the assessment of peak locations and their relative amplitudes [26,27], (i) magnetic resonance spectroscopy (MRS) measurements [14,17]; (ii) multi-echo Spoiled Gradient Recalled (SPGR) sequences [17,26,28] where it is assumed that the peak frequencies are known and do not change voxel by voxel while the variables related to the peak amplitudes are uncorrelated [17] or have to satisfy certain constraints imposed by the lipid chemical composition [29]; and (iii) multi-echo gradient-echo sequences with a high number of echoes [30]. All of those approaches are based on the hypothesis that fat spectral properties are spatial invariant. While this approximation for the frequency pecks is valid in almost most of the cases, in contrast, the lipid metabolites concentration (hence the amplitude of pecks) varies with the tissue, patient, anatomical site, composition and nature (i.e. metabolic or pathologic) of the fatty inclusions and this dependence is as large as in some pathologies linked to neuromuscular disorders (i.e. Duchenne, myopathies) progress [31–34]. Several researches demonstrated that accurate spectral modeling of fat is necessary when performing fat-corrected $R_2^*$ measurements [3,14,18,26]. Clinically, recent studies have demonstrated consistent differences in lipid compositions of adipose tissues in different anatomical regions [35] [36–38]. Particularly, applications on muscular fat in vivo [39,40] showed marked intra- and inter-individual variability of the spatial distribution of lipids in the musculature of the lower body. Similar conclusions have been also drawn by analyzing leg muscles in subjects affected by Duchenne muscular dystrophy (DMD) [7,41,42] or in the study of whole body fat distribution [43,44]. In addition, MRI has revealed patterns of selective muscle involvement in muscular pathologies that in a number of cases appear to be reasonably disease-specific [45,46], and the characterization of these patterns is necessary in order to improve the therapeutic perspectives of NMD patients. Finally, although most of such lipid accumulation in the liver and other organs usually does not exceed 50%, this is not in the case for degenerative muscle diseases, where higher FF values can be reached up to the complete substitution of the muscular tissue with fat and fibrosis [5,47–49] leaving the problem of the spatial variability of fat composition an important issue. According to this, the chance to perform a multi-regional self-calibration procedure on a given MRI dataset can be considered of interest in the study muscular dystrophies, being that the pathophysiological changes might spatially alter tissue relaxation properties [31,32] when evaluating affected from unaffected tissues. To this end, the signal model formulation with independent $R_2^*$ decay as provided in Ref. [50] has been taken into account in our study, being that it has been shown to be unsusceptible to some artifacts that adversely affect single-decay formulation when large regional variations of transverse decay rate occur among biological species [50]. This is the key motivation of this work where we have generalized the recent implementations of CSE water/fat separation techniques in skeletal muscle [20,49,51] that use a spatially constant pre-calibrated multi-peak fat spectrum model [14,17] by considering a spatially-variable spectrum model of the fat. We exploit a clusterization technique to segment a multi-echo NMR image dataset into a number of partitioned areas (PAs) under some chosen metrics. In other words, we develop an iterative self-calibration technique that for each PA enables the estimation of fat spectrum relative amplitudes to provide a PA-specific calibrated parameters and improve (i.e. reducing fitting error) overall CSE water/fat quantification process.

This algorithm is organized in the following steps (further details are provided in Section 3.2): (*i-ii*) segmentation of the input MRI space into a finite number of partitions by using the Gap-statistics, (*iii*) self-calibration on each partition, to extract corresponding fat spectral components, (*iv-v*) independent fat/water quantification on each partition using calibrated fat relative amplitudes, and consolidation of intermediate results and finally, (*vi*) generation of the output maps. The validation of the method has been performed by evaluating the numerical results as obtained from a cohort of enrolled patients, and analyzed in terms of accuracy, robustness and reproducibility. Accuracy is assessed by taking in consideration the Mean Squared Error (MSE) and cumulative MSE. Robustness has been evaluated by examining the occurrence of artifacts (i.e. fat-water swaps). Reproducibility has been investigated by analyzing multiple datasets from a subset of patients which agreed to undergo a supplemental examination after a 1-week period, and by calculating the Intra-Class Correlation (ICC). In the second part of the work the performance of the proposed approach are validated and compared with state-of-the-art CSE methods (in Section 4) focusing in the field of NMD and by considering ad-hoc and public MRI datasets. The structure of the paper is the following: Section 2 describes the key theoretical elements behind the proposed methodology. Then, in Section 3 we introduce the full reconstruction technique for fat/water quantification based on clustering and gap statistics and provide details about the subjects involved in the study and the reference metrics used to compare the results. Experimental findings are thoroughly presented in Section 4, whereas discussions are provided in Section 5. Finally, we draw conclusions in Section 6. Further results and explanations are included in the Supplementary Material and in the Appendix, respectively.

**2. Theory**

*2.1 Chemical Shift Methods without clusterization*

A CSE model for the fat/water quantification from a signal $s_q$ measured on a given voxel $q$ ($q=1,...,Q$, where $Q$ is the number of voxels) at time $TE_n$ ($n=1,...,N_{echo}$) is given by the following equation [17,19,26,30]:

$$s_q(t_n) = \left( \rho_{W,q} + \rho_{F,q} \sum_{p=1}^{P} \alpha_p e^{i2\pi f_{F,p} TE_n} \right) e^{i(\phi_{0,q} + 2\pi f_{B,q} TE_n)} e^{-R^*_{2C,q} TE_n} \qquad (1)$$

where $\rho_{W,q}$ and $\rho_{F,q}$ are the amplitudes of water and fat signals, respectively, with initial phase $\phi_{0,q}$, $f_{B,q}$ is the frequency shift due the spatial inhomogeneities of the bias magnetic field $B_0$. The terms $f_{F,p}$ are the known frequencies for the multiple spectral peaks of the fat signal relative to the water peak. Each fat peak $p$ has a different unknown amplitude $\alpha_p$ and $\sum_{p=1}^{P}\alpha_p = 1$ (being $P$ the number of fat peaks, here fixed as $P = 6$) and it is supposed that in Eq. (1) $f_{F,p}$ and $\alpha_p$ once defined, are spatially invariant. The usual assumption of a common relaxation rate (single-decay), $R_{2C,q}^*$, is used for both water and fat species. Although it has been recently documented how single-decay hypothesis can be subjected to numerical artifacts [50] under particular conditions, it is still widely adopted for its improved noise stability [52,53] and reproducibility in many traditional applications [15,16,20]. However, we highlight that the validity of the similarity assumption for water and fat signal decay rates is generally untrue as there is no physiologic basis for this assumption [19]. A refined variant is constituted by the following equation:

$$s_q(t_n) = \left( \rho_{W,q} e^{-R_{2W,q}^* TE_n} + \rho_{F,q} \sum_{p=1}^{P} \alpha_p e^{i2\pi f_{F,p} TE_n} e^{-R_{2F,q}^* TE_n} \right) e^{i(\phi_{0,q} + 2\pi f_{B,q} TE_n)} \quad (2)$$

Here, independent relaxation rates for water $R_{2W,q}^*$ and fat $R_{2F,q}^*$ are modeled [50]. Either signal models with single-decay approximation $R_{2C,q}^*$ as in Eq. (1) or independent decay ($R_{2W,q}^*$, $R_{2F,q}^*$) as in Eq. (2) can be calculated using Non-Linear Least Square method (NLLS) which provides the maximum-likelihood estimation [54] and are considered valid for each voxel. On a general basis, the relative amplitudes $\{\alpha_p\}_{p=1,\ldots,P}$ are among the unknowns to be estimated. In this research, we will consider the solution of Eq. (2) as the core of our self-calibration technique which will be described further.

*2.2 Clustering*

Clusterization is an important technique in data analysis and mining. The goal of clustering is to split a set of elements into subsets with some criteria of similarities. The elements of the same subset are more similar to each other than the elements from different subsets [55]. As the clustering problem requires an unsupervised approach, the definition of reliable parameters is a key ingredient in the segmentation process. There are many standard methods in the literature for clustering, among them, we cite hierarchical [56], partitioning [57–61], hybrid method [62], density-based [63] and fuzzy clustering [64–66]. Thanks to its manifold applications, recent efforts in the processing of MRI data are documented [67–70], including unsupervised assessment of fat distribution [71–75], segmentation [74,76–81] and whole-image optimization [82]. In Kullberg et al [74], a fully automated algorithm for segmentation of the visceral, subcutaneous, and total adipose tissue (TAT) depots from whole-body water and fat MRI data has been presented. In Berglund et al. [76] a simple segmentation of adipose tissue was performed, in order to quantify TAT, although such technique leverages on a traditional Three-Point Dixon [83,84] acquisition scheme, which generally provides limited information on relaxation maps if compared with modern multi-echo sequences [19,85,86]. Among several different clustering schemes, the $k$-Means [55] is a partitioning algorithm that, given a positive, finite and discrete number $k$, it is able to subdivide a dataset of $Q$ data points into $k$ groups (or clusters), by attempting to minimize the distance (we choose the Euclidean distance as the reference metric) between data points within a cluster and a point designated as the center of that cluster (intra-cluster distance). Here, $k$ data objects are randomly selected as centers (centroids) to represent $k$ clusters and remaining all data objects are placed in the cluster having center nearest to that centroid. After processing all data objects, new centroids are determined which can represent clusters in a better way and the entire process is repeated. In each new iteration, the position of centroids randomly changes until they minimize a sum of pairwise dissimilarities [58,60], and all data objects are bound to the clusters based on the new centroids. This process concludes when all the centroids do not move. As a result, $k$ clusters are found representing a set of input data objects, such that each object has the lowest intra-cluster distance (highest intra-cluster similarity) and the highest inter-cluster distance (lowest inter-cluster similarity), at the same time. A complete description of the algorithm is out of the scope of the paper and is provided in Ref. [87]. The $k$-Means is a favorable choice because it has demonstrated to be more computationally efficient than Fuzzy $C$-Means [66,88,89], Quality Threshold (QT) [61], Diff-Fuzzy [65] and Density-Based Spatial Clustering [63] methods, respectively. Here, the role of the clusterization is instrumental to perform segmentation of the input MRI data and consequently to being able to execute multiple calibration routines on such partitions.

*2.3 Chemical Shift Methods after clusterization and multiple calibration procedures*

First of all, we subdivide the input MRI data space into a number of $\Gamma$ of non-overlapping partitions where to perform an independent self-calibration process by calculating the solution of the Eq. (2) on the voxels pertaining a given partition and having the highest available fat SNR. Such intermediate results are then spatially averaged and $\{\hat{\alpha}_{p,\gamma}\} \equiv \{\alpha_{p,\gamma}\}_{\substack{p=1,\ldots,P \\ \gamma=1,\ldots,\Gamma}}$ can be extracted. Once collected, such parameters are used to feed the subsequent analysis step that will be performed over the entire MRI array. In this case, we will use the independent $\{\hat{\alpha}_{p,\gamma}\}$ estimations for each PA to determine fat/water composition, therefore the Eq. (2) becomes:

$$s_q(t_n) \underset{q \in V_\gamma}{\cong} \left( \rho_{W,q} e^{-R^*_{2W,q}TE_n} + \rho_{F,q} \sum_{p=1}^{P} \hat{\alpha}_{p,\gamma} e^{i2\pi f_{F,p}TE_n} e^{-R^*_{2F,q}TE_n} \right) e^{i(\phi_{0,q}+2\pi f_{B,q}TE_n)} \quad (3)$$

where $s_q$ is the signal at a time $TE_n$ of a voxel $q$ into the $\gamma$-th PA (whereas $q \in V_\gamma$, and $V_\gamma$ is the subset of the voxels contained in this PA, such that $\bigcup_{\gamma=1}^{\Gamma} V_\gamma = Q$). Therefore, the main difference that arises between Eq. (2) and Eq. (3) is that the fat relative amplitudes are unknowns in Eq. (2) whereas have been estimated in Eq. (3) and therefore they simplify the computational problem.

*2.4 Gap Statistics*

A fundamental problem in cluster analysis is determining an advantageous number of groups (or clusters) $\Gamma$ in a dataset [55], and a variety of more or less successful methods have been suggested to accomplish this goal [90–92]. When a clustering problem is posed as an optimization problem, the standard way to proceed is to try to optimize some cluster validity metrics [93] which can be internal measures, such as the ratio of within-cluster and between cluster similarities, or might follow information-theoretic criteria [94,95], stability [96,97] and statistical approaches [98], respectively. Indeed, some of such internal measures can be efficiently used to estimate the most suitable number of clusters in a dataset. It usually involves the computation of clustering results for a range of different numbers of groups, $M$, and the subsequent performance analysis under the chosen internal measure as a function of $i$ $(i \in M)$. Once completed, the method aims to determine a number $\Gamma$ (such that $1 \leq \Gamma \leq M$) of partitions to represent the MR image dataset. By calculating the gap function $\{G_i\}$, we can select the number of clusters $\Gamma$ to be the one that gives the maximum gap $G_\Gamma$. Details on how to perform the gap statistics and derive $\Gamma$ are provided in the Appendix. Thanks to the gap statistics we are able to identify a favorable number of groups to be used for the clusterization of our input dataset. These PAs are such that corresponding voxels have both maximum intra-cluster similarity (i.e. we consider the Euclidean distance between a voxel and its closest centroid) and inter-cluster diversity (i.e. we consider the Euclidean distance between a voxel and the other centroids). For a given volume slice a complex (both magnitude and phase) multi-echo sequence is extracted from MRI data and used as input.

## 3. Methods
*3.1 Our Gap-Driven Self-Calibration Algorithm*

Once complex MRI data are extracted, they are arranged into 5-dimensional matrices that, in its mostly used format [99] are composed by $N_x \cdot N_y \cdot N_z \cdot N_c \cdot N_{echo}$, where $N_x$, $N_y$, $N_z$ are the number of voxels in the 3-dimensional space. In order to simplify the concept, we will assume that the multi-coil data $N_c$ are combined together according to Ref. [100], so that $N_c = 1$. In our study, we will primarily focus on a single slice MRI data arrays ($N_z = 1$), therefore we have $\left[(N_x \cdot N_y) \times N_{echo}\right]$ array data elements as input objects for clusterization. The extension to multiple slices is straightforward. By considering the original implementation of the Mixed-Fitting method [16], the main differences in the Eq. (2) in terms of specifications are detailed below:

- Six-peaks fat model with the following relative frequencies (expressed in ppm) of (0.6, -0.5, -1.95, -2.6, -3.4, -3.8) [2,30,101–103];
- Independent estimations for water and fat decay rates ($R^*_{2W}, R^*_{2F}$) [19];
- Different variation ranges for water and fat relaxation rates, $R^*_{2W}$ [0-1000] s$^{-1}$ and $R^*_{2F}$ [0-200] s$^{-1}$, respectively, according to clinical evidence [31,104–106];
- Independent estimations of the relative amplitudes of fat peaks $\{\alpha_p\}_{p=1,..,P}$, (*p*=1,...6) [17];

To this aim, we use an oversampled multi-echo acquisition sequence ($N_{echo} \geq 15$) to collect MRI data as rapidly as possible in a single shot.

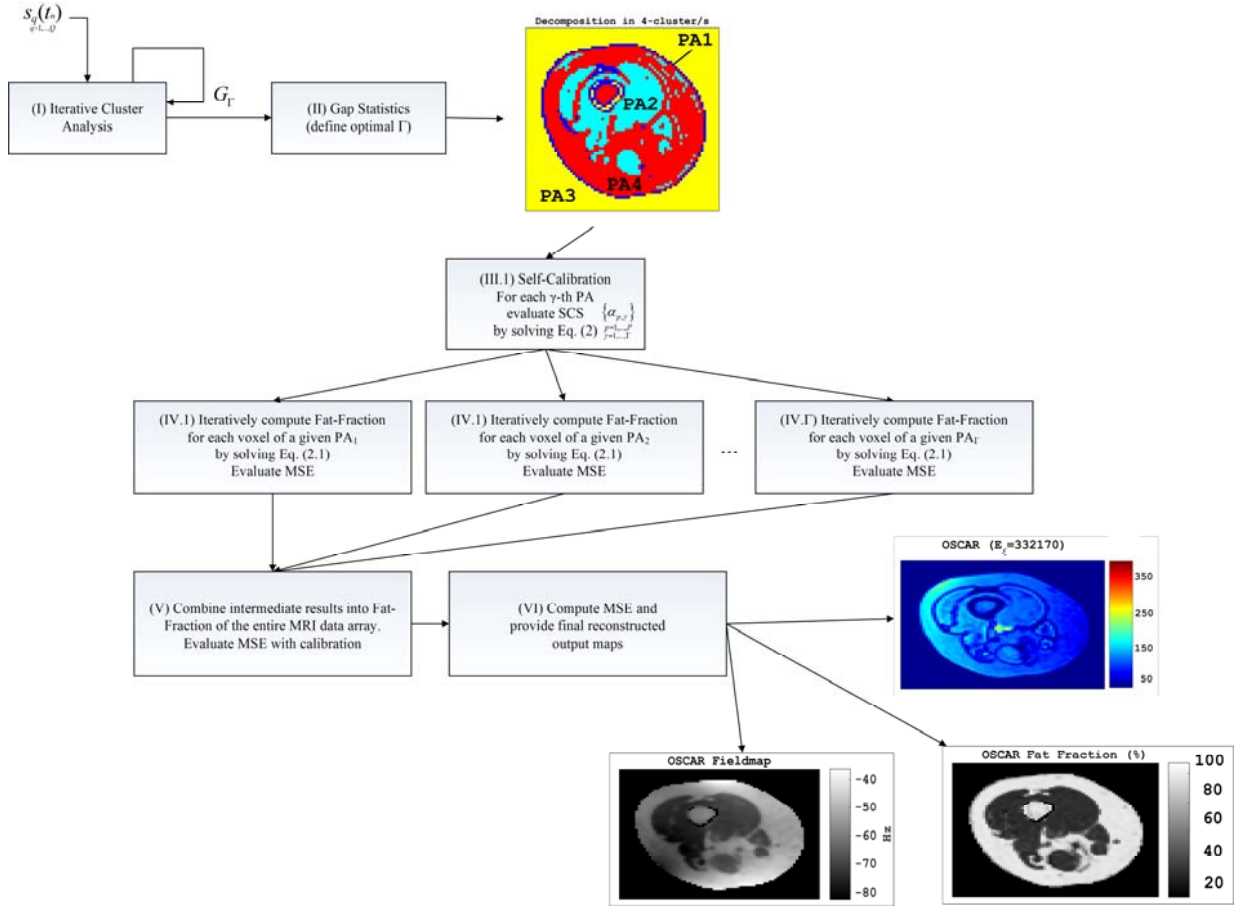

*** *Figure* 1 HERE ***

*3.2 OSCAR flow diagram*

Fig. 1 illustrates the flow diagram of the proposed method. Starting from an input multi-echo MRI dataset $s_q(t_n)$ (where $q=1,...,Q$), each block on the diagram performs different I/O operations and it is numbered depending on whether it is used to provide preparatory (sequential) or complementary (parallel) computations. The role of each computational block is summarized below: a raw multi-echo 2D SPGR sequence is processed in (I) using an iterative cluster analysis. Such analysis is carried out by considering a number *i* of possible cluster groups within the range $[1 \div M]$. Then, gap statistics is applied (II). Here, at each *i*-th step the corresponding dispersion functions are computed from original $D_i$ and from reference data sets $\overline{D}_{i,j}$ (where $j=1,...,L$ as detailed in the Appendix) in order to calculate the gap $G_i$. At the end of the analysis we are able to automatically obtain $G_\Gamma$ and therefore all the resulting PAs. For a given PA a representative

ROI is defined and will be hereinafter used to calculate (III.1) the corresponding *self-calibration subset* (SCS), $\{\hat{\alpha}_{p,\gamma}\}$, by solving Eq. (2). Once all the SCSs have been calculated, a systematic voxel-wise analysis (IV) is performed on all the elements of a given PA by solving Eq. (3) using a variant [50] of the Mixed-Fitting method in order to provide independent $R_2^*$-correction and $B_0$ field map estimation [15]. Such intermediate results are combined together (V) to consolidate the output data and obtain final reconstructed PDFF, fitting error and field maps. At last, (VI) MSE is calculated on the output by analyzing the fitting error between predicted results with calibration and experimental measurements. The smaller the MSE is achieved using self-calibration the larger is the accuracy in PDFF quantification. On an AMD Opteron-based server with 128 GB of random-access memory and four 2.2 GHz multi-core central processing units, solving this problem requires less than 20 minutes for each of the dataset analyzed.

*3.3 Performance metrics*

Above presented iterative approach was pursued to achieve the least squared fitting error $\xi$ [107] between acquire MRI data and the signal model. Given a number $\Gamma$ of clusters by means of gap statistics, for each cluster $\gamma$ ($\gamma=1,...,\Gamma$) we attempt to numerically solve the Eq. (2) in order to estimate the fat spectrum relative amplitudes $\{\hat{\alpha}_{p,\gamma}\}$. Above group of parameters will be hereinafter named *self-calibration subset* (SCS). Once such subsets are calculated for every segmented area, we are able to start a second iterative process that systematically performs Complex Multi-Echo $T_2^*$-corrected algorithm [30,50,102,103,108,109] on the voxels of each PA by using available SCS data. Here, for a given PA $\gamma$, the algorithm computes Eq. (3) over all the voxels, using available $\{\hat{\alpha}_{p,\gamma}\}$. No a-priori fat models (according to the ones analyzed in Ref. [102]) are used in this study. In order to show the advantages in terms of accuracy, robustness and reproducibility, different criteria have been taken into account. Accuracy is assessed by considering the Mean Square Error (MSE), $\xi$, as a standard measure [50] to quantify the fitting performance, which can be defined, for a voxel $q$, at a time $t_n$, as :

$$\xi_q \underset{q \in Q}{=} \min_{(\rho_{W,q}, \rho_{F,q}, R_{2W,q}^*, R_{2F,q}^*, f_{B,q}, \{\hat{\alpha}_{P,\gamma}\})} \sum_n |s_{q,est}(t_n) - s_q(t_n)|^2 \qquad (4)$$

where $s_{q,est}$ is the estimated signal as calculated using Eq. (3), and the cumulative MSE, $E_\xi$, defined as :

$$\mathrm{E}_\xi = \frac{1}{Q}\sum_{q=1}^{Q}\xi_q \qquad (5)$$

according to Ref. [110] and compare the results against other CSE algorithms by evaluating both field and PDFF maps (included as Supplemental Materials). As cited above, robustness has been evaluated by examining the occurrence of artifacts (i.e. fat-water swaps) [82,85,111] in the results due to low SNR regions or large field inhomogeneities [15] and by comparing them with other state-of-the-art methods. These localized swaps represent estimation errors where the main signal component in a voxel is assigned to the wrong chemical species (e.g., identifying as mostly water a voxel that contains mostly fat, or vice versa). On the other hand, reproducibility has been investigated by analyzing multiple datasets from a part of the cohort of patients which agreed to undergo a supplemental examination after 1 week, and by comparing the results as obtained from both tests. The correspondence between the estimated sets of $\{\hat{\alpha}_{p,\gamma}\}$ and the fat spectral distribution has been evaluated using MRS and biopsy [51] as have been collected in subcutaneous fat, muscular districts from healthy subjects and from patients having pathological adipose infiltrations (i.e. Duchenne [4,112] and other NMDs [50,51]), whereas the combined effects of multiple self-calibration routines with a $T_2^*$-corrected CSE model will be analyzed on fat quantification with applications on the screening, monitoring and disease assessment of neuromuscular disorders. Field-map smoothing [76,113] and regularization [82,114] techniques have been also considered in this study and Graph-Cut algorithm [15] has been used in order to provide stable initial field-map estimations. On the other hand, no additional temperature effects [25] and susceptibility effects [24] have been taken into account.

*3.4 Datasets Used for the Experiments*

The proposed method has been tested on 24 subjects ([17 male, 7 female]; age 31.2 ± 18.4 y [range, 7 - 68 y]), of which 3 were healthy volunteers whereas 21 where subjects with moderate (5 partial loss of functional but still ambulant), mild (9 intermittently assisted by wheelchair or other similar mobility devices) and severe (7 completely wheelchair-dependent) functional impairment, and confirmed diagnosis of NMD. Subjects underwent examinations in the supine position with a standard multi-channel phased-array coil on Philips Achieva 1.5 T, 3 T and Ingenia 1.5 T scanners (Philips, Best, The Netherlands), in accordance with the local institutional review board. MR images were obtained by performing a multi-echo 2D SPGR acquisition with "fly-back" gradients. Candidates were recruited at the Neuromuscular Disorders Center of the University of Messina with definitive diagnoses of DMD, inclusion-body myositis (IBM), Facioscapulohumeral disease (FSHD), Pompe and McArdle disease, and underwent MR imaging on the pelvis and upper leg muscles for the disease clinical assessment between 2009 and 2016 and had been partly published in previous works [50,51,115]. Inclusion criteria were: confirmed NMD diagnosis and no severe or moderate learning difficulties or behavioral problems. Exclusion criteria were: contraindications to MRI, and inability to cooperate and participate in the various tests. These patients had no history of chronic illness other than NMD (including any neuromuscular, metabolic, or endocrine disorder that could alter bone or muscle metabolism). The multi-slice coil-combined datasets correspond to similar anatomical regions (i.e. thigh and pelvis). The masks were generated by thresholding the raw images and validated by an expert radiologist. A subset of subjects was re-examined after a 1-week period in order to evaluate reproducibility of the method in PDFF quantification. OSCAR has been also tested on a publicly available abdominal dataset for reference purposes. Subjects characteristics of the cohort are summarized in Table 1.

| Disease Type | Gender | Age (Range) | Disease severity | FF Range | Functional mobility | Dataset Type |
|---|---|---|---|---|---|---|
| Duchenne muscular dystrophy (DMD) | 5 (all male) | 8.40±1.51y (7-10y) | Moderate to Medium | 15-30% | 2 amb, 3 wcd | Thigh |
| Inclusion-body myositis (IBM) | 1 (male) | 27y (27y) | Medium | ~30% | 1 amb | Thigh |
| Pompe disease | 6 (3 male, 3 female) | 39.0±17.4y (13-62y) | Moderate to Medium | 15-30% | 4 amb, 2 wcd | Pelvis, Thigh |
| Limb Girdle disease | 1 (female) | 45y (45y) | Medium | 35% | 1 amb | Pelvis, Thigh |
| Becker muscular dystrophy | 3 (3 male) | 21±9.53y (15-32y) | Moderate to Medium | 15-30% | 3 amb | Thigh |
| Facioscapulohumeral muscular dystrophy (FSHD) | 5 (male) | 48.6±15.8y (26-68y) | Moderate to Severe | 25-70% | 3 amb, 2 wcd | Thigh |

| Healthy volunteers | 3 (male) | 32±7.5y (25-40) | None | 1-10% | 3 amb | Thigh |
| --- | --- | --- | --- | --- | --- | --- |
| **Total** | **24** | **31.2±18.4y (7-68)** | **None to Severe** | **1-70%** | **17 amb, 7 wcd** | **Pelvis, Thigh** |

*\*\*\* Table* 1 *HERE \*\*\**

An unipolar multiecho SPGR acquisition was performed with the following typical parameters: $TE_1/\Delta TE$ = 1.6/1.3 ms, FOV = 40x40cm, reconstruction matrix size = 256x256, flip angle=5°, repetition time (TR) = 60 ms. In this study, the multipeak self-calibration for in vivo fitting was performed on a 16-echo in vivo dataset. For reference purposes (see Section 4.5), the $T_1$-weighted Fast Spin Echo (FSE) MRI data were obtained with a flip angle = 90°, reconstruction matrix size = 640x640, $TE_1$ = 10 ms, TR = 400 ms. In order to compare the method with a publicly available pelvis dataset [99], a minimum of 15-echoes has been taken into account. By visual inspection of the resulting decompositions, we focused our attention on the scenarios containing areas with relevant fitting errors.

## 4. Results

*4.1 Example using a Thigh MRI dataset*

Fig. 2 summarizes the results of a reference 1.5 T thigh MRI dataset from an NMD subject with medium severity of the disease, which has been also included in the reproducibility study (see Subject #4 in Fig. 7). Fig. 2 (a) displays the evaluation of the Gap function $G_\Gamma$ and the sum of intra and inter-cluster distances (left and right insets of Fig. 2 (a), respectively). In order to obtain an useful clustering of the input data set, we should select a large-enough cluster range $1 \leq i \leq M$, such it is possible to maximize the gap statistics function. According to Ref. [98], we choose the cluster size $\Gamma$, to be the smallest number $i$, such that $G_i \geq G_{i+1}$ - $s_{i+1}$, which identifies the point at which the rate of increase of the gap statistic begins to slow down. In this case, $\Gamma$ = 5, which corresponds to a trade-off between minimum intra-cluster distance (left inset of Fig. 2 (a)) and maximum intra-cluster distance (right inset of Fig. 2 (a)). Reference magnitude image of the input MRI dataset is provided in (b) whereas in (c) we observe the result of the corresponding segmentation. Interestingly, not all the PAs are topologically connected (in particular PA1, PA2), this is a desired property of the clusterization algorithm which is able to assign elements to a given group also if they are not spatially contiguous. PA1 (blue colored areas) is related to the muscular groups which are minimally or still not affected by the disease (i.e. sparse or limited fatty infiltrations can be seen from the related $T_1$-weighted FSE images), PA2 (cyan colored areas) collects most of the voxels with pathological lipid infiltrations (again, such correspondence can be seen by evaluating $T_1$-weighted FSE images for reference purposes), whereas PA3 (green colored areas) is mainly related to common adipose tissue.

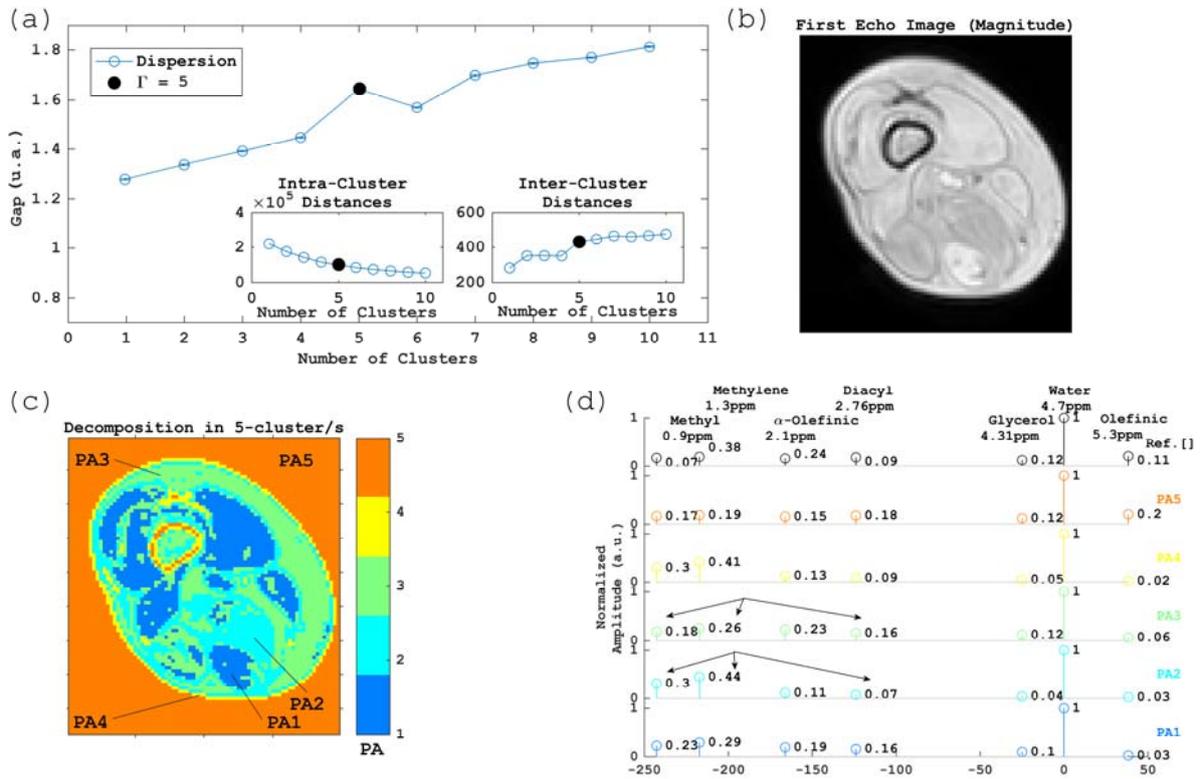

*** *Figure* 2 HERE ***

In Fig. 2 (d) the summary of the resulting spectral fat distribution on each PA is reported together with a reference computation by using the method described by Yu et al. [17] (black bars) which adopts a single self-calibration routine. We can observe a remarkable difference in fat spectral composition by evaluating the fat relative amplitudes between PA2 and PA3 which account for different fat deposits. In PA2 the large majority of voxels pertains to sites with significant pathological infiltrations. In such regions, the most meaningful components of the fat signal are Methyl {-(CH$_2$)$_n$-CH$_3$} and Methylene {-(CH$_2$)$_n$-} (0.3 and 0.44, respectively). On the other hand, in PA3, where the presence of fat in voxels has a generally physiological origin, together with Methyl and Methylene groups, also the α-Olefinic {–CH$_2$–CH=CH–CH$_2$–} is significant. Besides, similar fat spectral components can be observed in PA1 with muscular tissues still unaffected by the disease, where, intuitively, the presence of fat infiltrations can be substantially ascribed to natural metabolic phenomena. Our results suggest that such heterogeneity in fat chemical composition among different areas can be exploited to increase fitting accuracy and by doing so we take advantage of the independent estimation among multiple partitions. It is worth to mention that, similarly to Yu et al. [17], this spectrum self-calibration approach is based on the assumption that all fat-containing pixels in a dataset can be characterized by the same spectral frequencies of the fat peaks, i.e., $f_{F,p}$ are considered spatially invariant. On the other hand, its key difference is that it tries to address the spatial variability of fat spectral relative amplitudes using a multiple self-calibration approach. Indeed, from our results we can observe how the relative amplitudes exhibit a diverse combination in different sites, and cannot be adequately represented by an unique global calibration process, as in many current approaches.

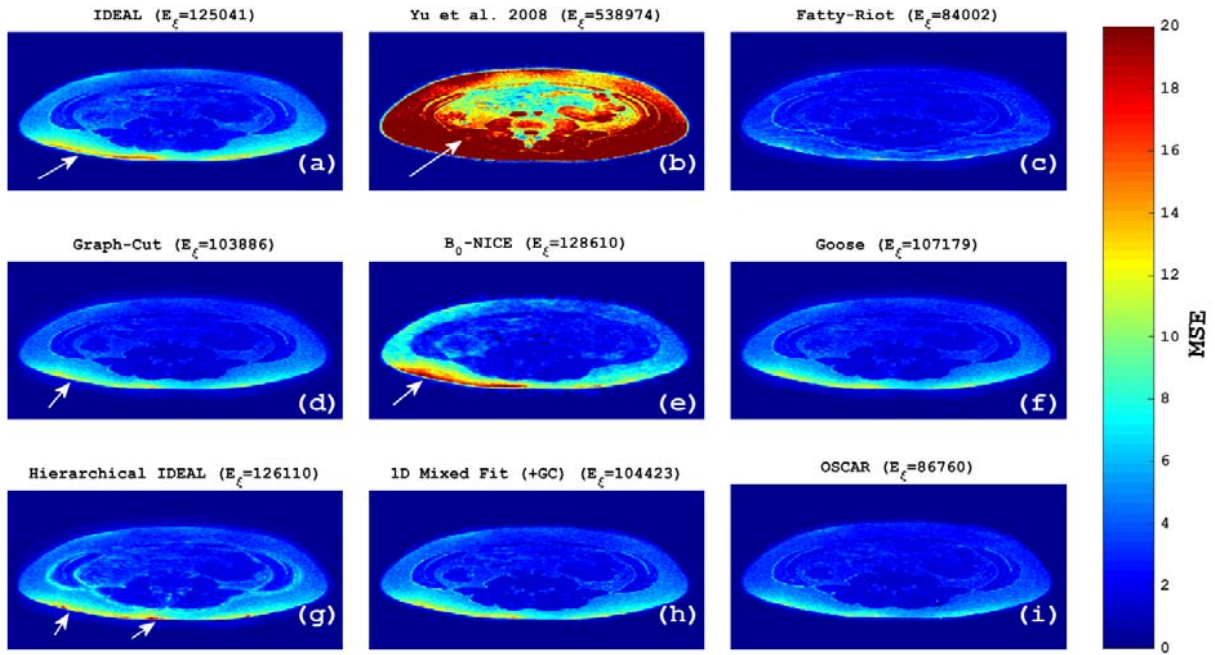

*** *Figure* 3 HERE ***

*4.2 Example using an Abdominal MRI dataset*

According to Fig. 3 (abdominal scan at 3 T, a 15-echoes dataset made available as part of an open challenge [99]), we evaluate the performance of OSCAR against several of the currently available techniques in terms of MSE, whereas the cumulative MSE ($E_\xi$) is provided in brackets. Specifically, a comparison is provided with a variety of state-of-the-art implementations (a-i): (a) IDEAL [13], (b) Yu et al. [17], (c) Berglund et al. [82], (d) Graph-Cut [15], (e) Non-iterative correction of phase-errors ($B_0$-NICE) in [116], (f) GlObally Optimal Surface Estimation (GOOSE) in [117], (g) Hierarchical IDEAL in [118], (h) Mixed Fitting with single $R_2^*$ decay (1D) approximation [16], and finally (i) OSCAR method. We provide also the resulting maps obtained for Fat Fraction and $B_0$ field inhomogeneity as supplemental materials Figs. 3.2-3.3, respectively. For each method, white arrows indicate the regions exhibiting significant fitting errors. It is worth noting how, among the presented methods, our approach (i) aims to achieve one of the lowest MSE, together with Graph-Cut (d). It is also interesting to observe from above cited supplemental materials how $B_0$ field does not show large inhomogeneities and PDFF maps are positively benefited by the adoption of a field smoothing function [119].

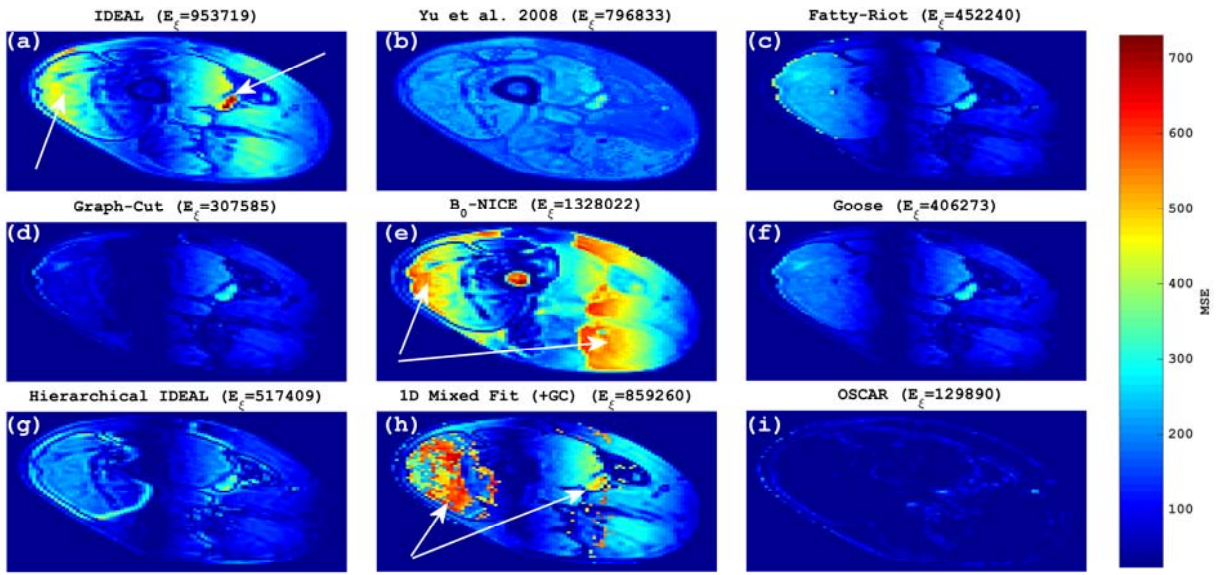

*** *Figure* 4 HERE ***

*4.3 Example using a Thigh MRI dataset of NMD patient*
In Fig. 4 (thigh scan at 1.5 T) we show the MSE results of a FSHD subject where a complete atrophy has been documented in thighs with medium disease severity and massive fatty substitution of the right semimembranosus, semitendinosus and gracilis muscle. The corresponding cumulative MSE is provided in brackets, whereas for the same methods a comparison is provided also in terms of PDFF and field maps in supplemental materials 4.2-4.3, respectively. A panel with the clusterization results is reported as supplementary material in Fig. 4.4. From the analysis of Fig. 4 we can argue how the most significant difference arises by comparing the IDEAL (a), $B_0$-NICE (d) and Mixed-Fitting (h) implementations where a large MSE is reported, which is probably related to the severe $B_0$ field fluctuations (see Fig. 4.3) which cause a decline in estimation accuracy, especially along the regions affected by such fluctuations (indicated as white arrows). From Fig. 4 (i) it can be observed how OSCAR (i) method provides a low MSE and also how fat-dominated tissues exhibit a generally reduced error if compared to the other approaches. Fat fraction maps (Fig. 4.2) offer useful insights to the evaluation of the different methods. Particularly, Yu et al. (b) shows an extensive water-swap region near the adductor muscle compartment. Methods (d, e, f) exhibit similar performance, whereas $B_0$-Nice (e) and Mixed-Fitting (h) reveal small artifacts along the areas affected by the largest fitting errors. From the analysis of field maps (see Fig. 4.3), we can observe that large localized $B_0$ fluctuations are present at the outer boundaries near the edges of the FOV. At the same time, it seems that both Berglund et al. (c), Hierarchical IDEAL (g), Mixed-Fitting (h) overestimate such field inhomogeneity by causing a corresponding increase of MSE.

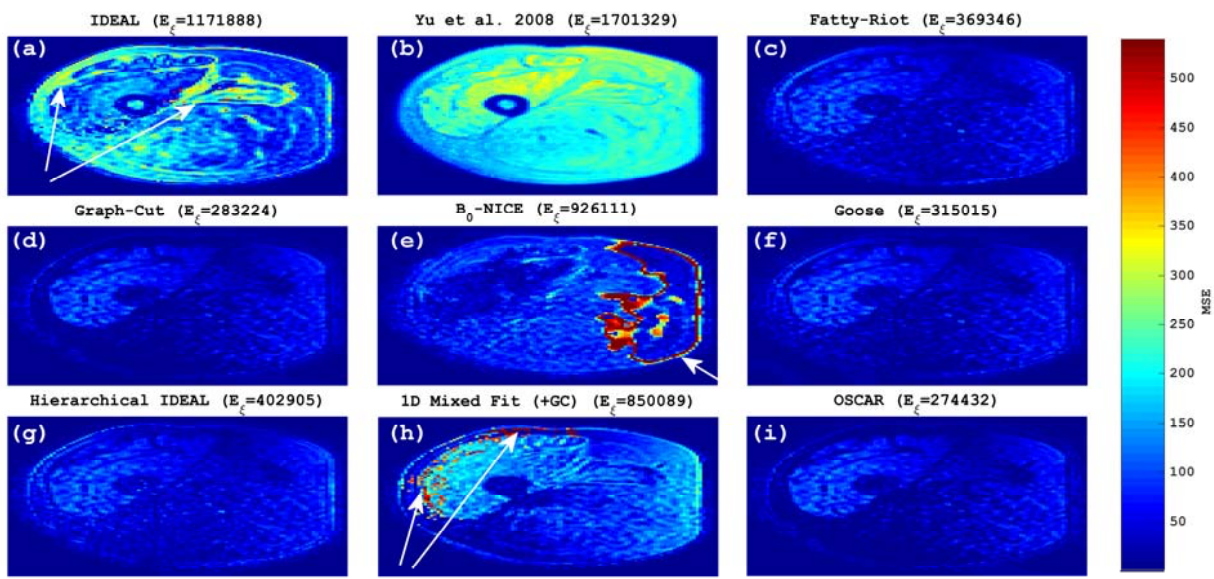

*** *Figure* 5 HERE ***

*4.4 Example using a Thigh MRI dataset of Pompe patients*

For the case of Fig. 5 (thigh scan at 1.5 T) we show the MSE results of another patient with known diagnosis of FSHD, having a substantial fat transformation of the thigh muscles. Disease severity is so significant that the majority of the muscular tissues are affected by pathological adipose infiltrations. Again, the corresponding cumulative MSE is provided in brackets. In this case, among the techniques under investigation, the approach proposed by Yu et al. [17] (b) provides the largest $E_\xi$, whereas in the Graph-Cut (d) and OSCAR (i) methods we observe the lowest fitting error, followed by Berglund et al. (c) and GOOSE (f).

To quantify the differences among the methods under test, the corresponding PDFF and inhomogeneity field maps are included in Supplementary Materials as Figures 5.2 and 5.3, respectively, whereas the panel with clusterization results in shown in Fig. 5.4. Interestingly, $B_0$-Nice (e), IDEAL (a) and Mixed-Fitting (h) exhibit different regions (e.g. subcutaneous adipose infiltrations near the sites with muscular tissues not completely affected) with larger MSE (as indicated by the white arrows in Fig. 5) with correspond to relevant estimation complications in the related PDFF map (see Fig. 5.2). The results indicate that methods in (c),(e),(f) and (i) yield qualitatively expected behavior in terms of fat fraction quantification, whereas (a), (d) and (h) a variety of fat/water swaps appear in the areas where an incorrect estimation of $B_0$ field inhomogeneity occurs (as indicated by the white arrows).

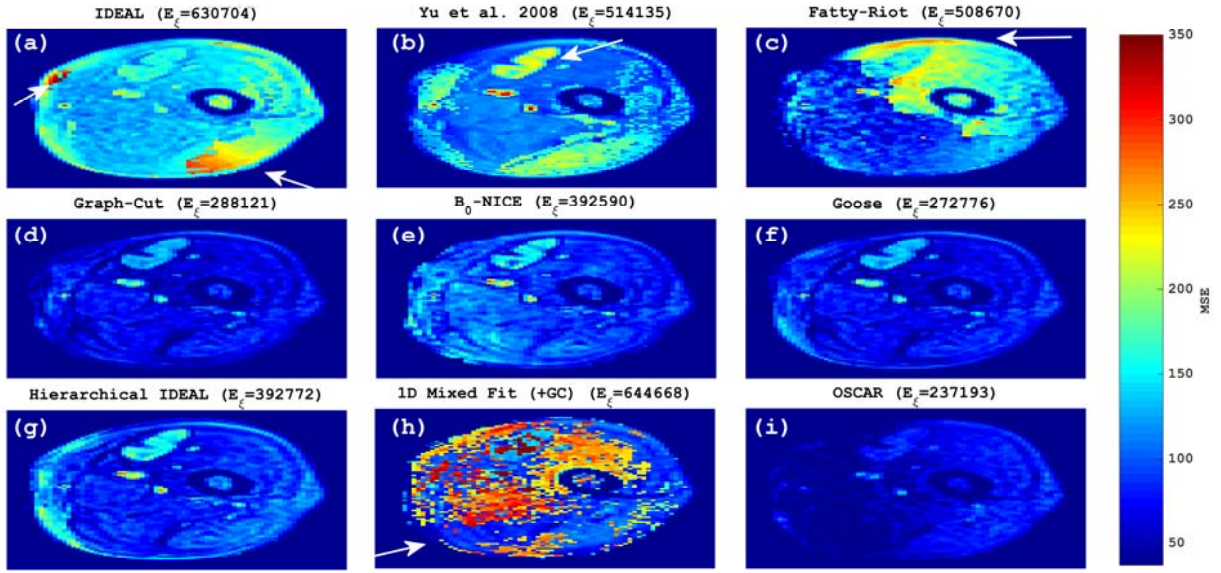

*** *Figure* 6 HERE ***

For the case of Fig. 6 (thigh scan at 1.5 T) we show the MSE results of a FSHD patient with severe disease conditions, caused by a complete atrophy and fatty transformation of the thigh muscles. The left rectus femoris muscle is spared. Hypointense thin lines surrounding involved muscles correlates with thickened perymisium. The corresponding cumulative MSE is provided in brackets. In this case, among the techniques under investigation, the IDEAL (a), Yu et al. [17] (b), Berglund et al. (c) and Mixed-Fitting provide the largest cumulative MSE, whereas Graph-Cut (d), Goose (f) together with OSCAR (i) exhibit the lowest fitting error, followed by methods (e) and (g). Again, the corresponding PDFF, inhomogeneity field maps and clusterization results are included in Supplementary Materials as Fig. 6.2-6.4, respectively.

We observe how IDEAL (a), Berglund et al. (c) and Mixed-Fitting (h) depict different regions with large estimation errors with corresponding problems in the related PDFF map (see Fig. 6.2) mainly because of the inaccuracy in quantifying $B_0$ field inhomogeneity.

*4.5 Reproducibility Study*

The reproducibility of one technology refers to the proximity in agreement between a series of measurements obtained from the same subject scanned under different conditions. Therefore, the reproducibility of the proposed technique to quantify fat-fraction is essential to ensure that the quantified data can be pooled from different time scales. The current prospective pilot study was conducted to assess the reproducibility of PDFF as a biomarker for muscular fat concentration in vivo, which aimed to evaluate the reproducibility of the method at different time scales, for different muscular groups of the thigh, and under diverse levels of disease progression. Such study was conducted for six subjects of the cohort under test that accepted for a second scan after a 1-week period in order to ensure a substantially unchanged health status. Different muscles have been taken into account by considering anterior muscular compartment (including vastus lateralis (VL) and vastus medialis (VM)) and posterior muscular compartment (including adductor (AD), biceps femoris (BF), semimembranosus (SM), semitendinosus (ST)). In Fig. 7 we report some representative transverse PDFF imaging (in units of %) results obtained from six volunteers during two tests on a 1-week interval. A $T_1$-weighted FSE MRI data (first column) is provided for reference purposes with the indications of above muscles, whereas the upper leg PDFF maps as obtained from the first and second scan (second and third column, respectively) were arranged and quantitatively compared. MRI-determined PDFF parametric maps of such subjects, demonstrating direct comparison of muscular fat content for almost six different ROIs placed on each leg during the tests. As can be observed from the results of NMD candidates ranging between moderate (Subjects 1 and 4), medium (Subjects 2 and 5) and severe (Subjects 3 and 6) lipid infiltrations in thigh muscles, the PDFF maps generally exhibit a close agreement. In Subject 3 (Severe) the progress of the disease is so relevant, that there is a dramatic, and almost total substitution of muscular tissue with fat and fibrosis on the left thigh, with both anterior and posterior compartments compromised. Only a very limited portion of left vastus medialis muscle is still recognizable. We observe how in this patient the PDFF map of second test underestimates the fat content along the vastus lateralis and biceps femoris muscles. We believe that this is due to the non-optimal positioning of the candidate and the occurrence of sharp local $B_0$ inhomogeneities during scan, more accurate results are obtained from first test. The close qualitative and quantitative agreements in MRI-determined PDFF were recorded. The parametric maps were generated from source images by applying OSCAR algorithm and are displayed with a PDFF ranging from 0%-100%. The main characteristics of the subjects under test are reported in Table 2.

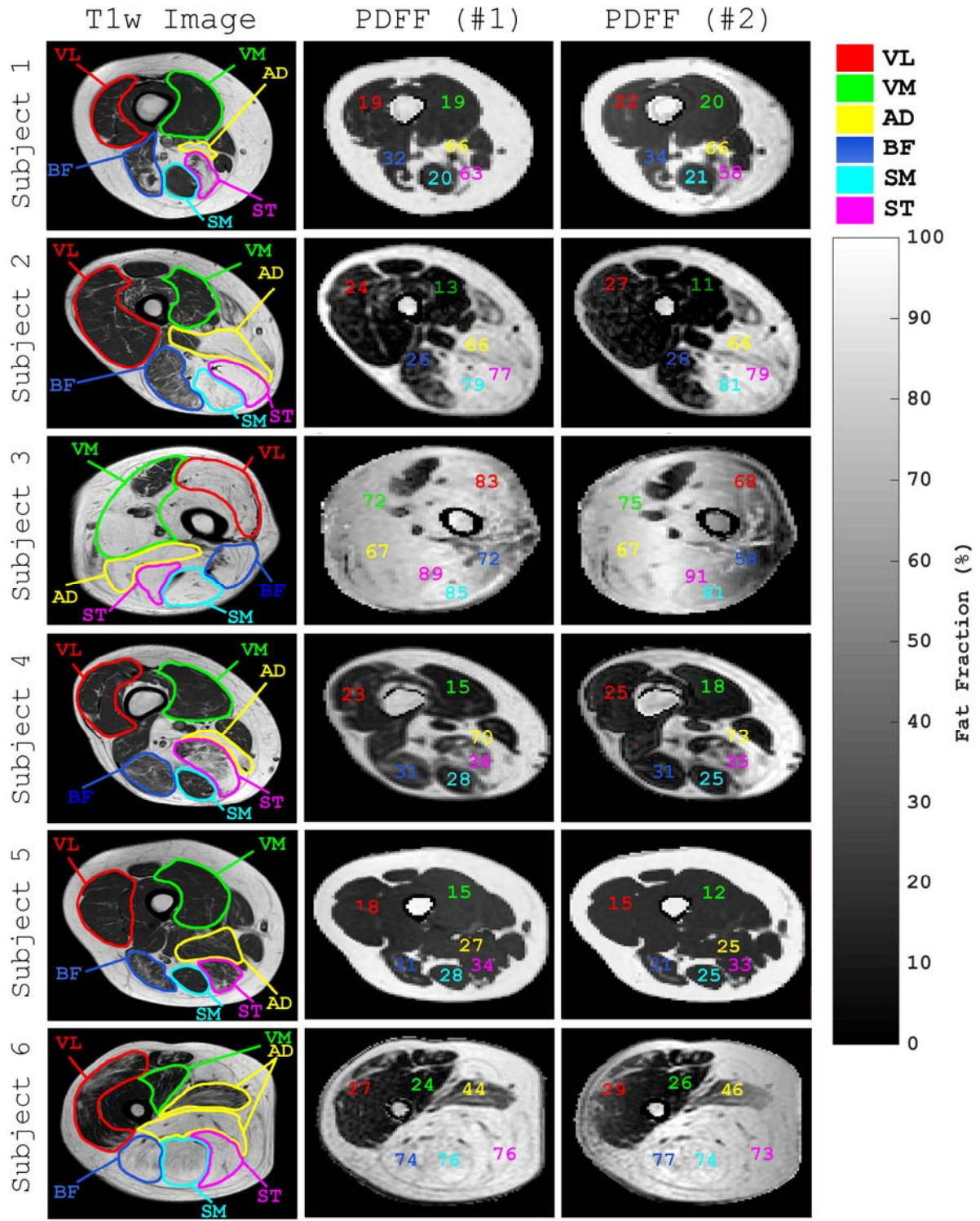

Reproducibility was measured by using the ICC coefficient [120]. ICC measures the contribution of inter-subject variances to total variance, which presents the ability of a method to detect the differences between subjects consistently. ICC values range from 0-1, and a value close to 1 indicates high reproducibility. Similarly to the study proposed in Ref. [121,122] we evaluated here ICC values which were calculated such as 95% confidence limits between two muscular PDFF measurement values. The ICCs of PDFFs related to upper leg muscles between first and second scan at one-week interval and across all tests were in the range of 0.893 and 0.977, respectively (Table 3).

|  | Description | Total ICC | 95% confidence interval | |
| --- | --- | --- | --- | --- |
| Subject 1 | 62-yo female Pompe (Moderate) | 0.969 | 0.957 | 0.977 |
| Subject 2 | 58-yo male FSHD (Medium) | 0.969 | 0.960 | 0.977 |
| Subject 3 | 68-yo male FSHD (Severe) | 0.893 | 0.861 | 0.918 |
| Subject 4 | 48-yo male FSHD (Moderate) | 0.958 | 0.944 | 0.968 |
| Subject 5 | 43-yo male FSHD (Medium) | 0.977 | 0.971 | 0.983 |
| Subject 6 | 68-yo male FSHD (Severe) | 0.919 | 0.901 | 0.934 |

*** Table 2 HERE ***

According to Fig.7 by evaluating the muscular PDFF in the ROIs, we have taken into account similar muscular compartments having a PDFF spanning along a broad range of values. Among candidates, three subjects exhibited moderate disease severity, two exhibited medium and one exhibited critical disease severity. Concerning the intra-scanner ICC PDFF, the 95% confidence limit was smaller for the PDFFs related to medium to severe disease (Subject 3 and 6 having ICC of 0.893 and 0.919, respectively) than when related to moderate disease (Subjects 1, 4 and 5 with ICC of 0.969, 0.958 and 0.977, respectively), which indicated that the reproducibility was slightly reduced as the disease severity increased (Fig. 7). Our results point out a low intra-subject variability in measurements for moderate to medium severity, which slightly increases when moving to critical severity.

## 5. Discussion

To develop a tool able to address spatial variability of fat spectral composition, the potential of clustering with gap statistics [98] in MRI data has been explored. Multiple self-calibration processes can yield more accurate fitting performance if compared to methods with single or no-calibration. Considering this, we carry out a multiple self-calibration procedure to obtain a set of fat relative peak amplitudes $\{\hat{\alpha}_{p,\gamma}\}$ on each of the $\gamma$ partitions, which hereinafter have been kept constant for the voxels of each given PA. Besides, in order to reduce the sensitivity of the reconstruction to possible confounding effects, a very low flip angle (5°) has been used in our study to minimize $T_1$-weighting [115], and an independent $T_2^*$-correction has been considered for both water and fat species, according to clinical evidence. Experimental findings have been compared against a variety of latest state-of-the-art CSE methods using reference metrics. Particularly, by evaluating results in Fig. 4, 5 and 6, the adoption of a regularization technique in OSCAR, based on Ref. [15] was instrumental to the achieved results since it mitigates the occurrence of fat/water swap artifacts (as reported in Fig. 4.2, 5.2 and 6.2) due to locally large $B_0$ inhomogeneities (as reported in Fig. 4.3, 5.3 and 6.3). In particular, by evaluating the Fig. 4.2, 5.2 and 6.2 the FF maps as obtained using the OSCAR method are also consistent between all slices (results not shown) and our observations have been substantiated under different examinations (as can be shown in Fig. 7). OSCAR has been tested using a combination of both ad hoc data set with NMD subjects and a publicly available dataset, demonstrating to reduce the MSE if compared with other competitive methods. The comparison of the algorithm against several state-of-the-art fat water decomposition algorithms enables to highlight some further advantages. Firstly, the formulation of OSCAR fully exploits the theoretical principles which are behind the gap statistic enabling a useful segmentation of a given multi-echo image dataset and finally the numerical solution of the fitting problem over the input NMR image space using different calibration datasets. Secondly, such methodology is shown to be robust, and it allows the automatic segmentation of MRI images to be carried out with no or limited user intervention, providing useful clinical indices for the characterization of muscular fat distribution and disease progression being also less prone to intra or inter-observer variability [51]. Thirdly, inhomogeneity field maps exhibit a constrained variation of the field that positively affects the fat/water quantification both in terms of estimation accuracy and noise sensitivity. Here, the proposed technique was developed in MatLab(R) (The Mathworks, Natick, MA, USA). Regarding the comparison with the other approaches, when related implementations were available, we assumed the default settings for reference purposes. The differences among the CSE methods in terms of adoption of field map smoothing/regularization, $T_2^*$-correction and self-calibration support are summarized in Table 3, together with the corresponding references.

| Index | Method | $\{\alpha_p\}_{p=1,...,P}$ Self-Calibration | $B_0$ Field map regularization | $T_2^*$-correction | References |
|---|---|---|---|---|---|
| (a) | IDEAL | No (A priori known)†(1) | No | No | [13] |
| (b) | Yu et al. | Yes. Single†(3,6) | No | Single | [17] |
| (c) | Berglund et al. | No (A priori known)†(9) | Yes | Single | [82] |
| (d) | Graph-Cut | No (A priori known)†(6) | Yes | Single | [15] |
| (e) | $B_0$-NICE | No (A priori known)†(6) | Yes | Single | [116] |
| (f) | GOOSE | No (A priori known)†(6) | Yes | Single | [117] |
| (g) | Hierarchical IDEAL | No (A priori known)†(6) | Yes | Single | [118] |
| (h) | Mixed Fitting with Graph-Cut initial estimates | No (A priori known)†(6) | Yes | Single | [16] |
| (i) | OSCAR | Yes. Multiple†(6) | Yes | Independent | This paper |

*** Table 3 HERE ***

All routines and studies were integrated and performed by means of a parallel processing framework which has been implemented for accelerating algorithms computation and recently demonstrated successful results in related [50] and other research fields [123–126]. It must be noted that, in addition to the methods compared in this article, there are several other recent methods that, for example, impose spatial constraints on the field map [127,128], or object-based information of the magnetic field inhomogeneity [129] to improve water/fat separation. A comparative study with these alternative methods is beyond the scope of this article. Several limitations exist in the current study. First, a larger number of severe NMD patients could not be included, as a 1-week re-examination period was too strict for most of them. Secondly, this study was performed on a small subset of healthy volunteers, moderate, mild and severe cases. Further studies with a larger sample of NMD patients should be conducted. Thirdly, multipeak fat modeling is limited to a six-peak model, even though more peaks can be found by spectroscopy [101].

## 6. Summary and Conclusions

This paper has presented a novel method for robust water/fat separation which has the ability to address the problem of the spatial variation of fat spectral composition due to intra- and inter-individual variability [40]. The proposed method uses a statistically motivated formulation to solve the fat/water quantification problem by subdividing the input MRI data set into a finite number of partitions via clusterization, performing self-calibration and finally calculate by means of a $T_2^*$-corrected multi-peak fat signal model.

Results established that the presented algorithm was found to perform robustly in NMD imaging studies with a generally better overall performance (average reduction of not less than 20% of fitting error, in terms of cumulative MSE) against latest CSE techniques that do or do not natively rely on preliminary calibration routines. Particularly, the PDFF of the thigh was more reproducible for the quantitative estimation of pathological muscular fat infiltrations, which may be applicable to evaluate disease progression in clinical practice. This technique is built on a general basis which makes possible further extensions of the proposed approach to several others CSE methods. Since PDFF quantification is now considered as a necessary measure to perform in MRI studies of all the patients with degenerative muscular diseases [130], we believe that the approach described here is particularly useful for the clinical applications where spatial differences in fat spectral composition currently prevent reliable water/fat separation.

## Acknowledgements

The simulations have been performed with the availability of SCL (Scientific Computing Laboratory) of the University of Messina, Italy. G. S. acknowledges project entitled "Tecniche innovative di processamento di segnali per lo sviluppo di sistemi e servizi ICT". This research did not receive any specific grant from funding agencies in the public, commercial, or not-for-profit sectors. The authors thank Prof. Junmin Liu for providing technical support.

**Appendix: Computational procedure of the gap statistic method**

Particularly, the gap statistic [98] is characterized by an established approach which can be briefly described as follows:

(*i*) For a given clusterization method, we define a maximum number of clusters $M$ in order that the value $\Gamma$ satisfies the condition $1 \leq \Gamma \leq M$ (we set $M = 50$);

(*ii*) Iterate the same procedure for each $i$ where ($i = 1,...,M$);

(*iii*) Run the *k*-Means clustering algorithm on the MRI dataset to find $i$ clusters, and sum the distance of all points from their cluster mean. Call this sum the dispersion $D_i$. Being that, for a given number of clusters, $i$, data are partitioned into $i$ clusters, we can define with $C_r$ the index of observations in a given cluster $r$ ($r = 1,...,i$) and $n_r = |C_r|$. Then, we can express dispersion as the following:

$$D_i = \sum_{r=1}^{i} \frac{1}{2n_r} \delta_r \quad (6)$$

where $\delta_r$ is the sum of the pairwise distances (again, we use Euclidean distance) for all objects $(p, \dot{p})$ in cluster $r$:

$$\delta_r = \sum_{p, \dot{p} \in C_r} r_{p\dot{p}} \quad (7)$$

(*iv*) Generate a set of $L$ reference datasets having the same size as the original. In this study, we set $L = 100$;

(*v*) Calculate the dispersion $\bar{D}_{i,j}$ ($j = 1,...,L$) of each of these reference datasets, and compute the mean;

(*vi*) Compute the *i*-th gap $G_i$ defined as:

$$G_i = \log\left(\frac{1}{L}\sum_{j=1}^{L} \bar{D}_{i,j}\right) - \log(D_i) \quad (8)$$

Once we have determined all the gaps $\{G_i\}$, we can select the number of clusters $\Gamma$ to be the one that gives the maximum gap $G_\Gamma$, therefore it can be expressed as:

$$G_\Gamma = \max\{G_i\}, \ (i = 1,...,M) \tag{9}$$

Briefly, the gap statistic [98] compares the change in within-cluster dispersion $D_i$ with that ($\bar{D}_{i,j}$) expected under an appropriate null distribution (i.e. the one iteratively obtained from the reference datasets in step (*iv*)) in order to find the maximum value which represents the proper partitioning.

**Figures captions**

**Fig. 1**. OSCAR flow diagram. [Color figure can be viewed in the online issue].

**Fig. 2**. (a) Characteristic of the Gap function $G_\Gamma$ using *k*-Means clusterization for a different number of clusters, *i*. Here the most favorable number of clusters is $\Gamma = 5$. (left inset of Fig. 2 (a)) Sum of intra-cluster distances as a function of *i* from the same clustering method. (right inset of Fig. 2 (a)) Sum of inter-cluster distances as a function of *i* from the same clustering method. (b) The magnitude for the first echo from the input data set. (c) The resulting cluster groups are not necessarily topologically connected (e.g. PA1, PA2). (d) Spectral information estimated using OSCAR (colored bars, one spectrum for each PA) and reference [17] pre-calibration approach (black bars). [Color figure can be viewed in the online issue].

**Fig. 3**. A comparison in terms of MSE for a public 3 T abdominal MRI dataset reported in Figure 4, among current implementation of most of state-of-the-art algorithms for Chemical Shift Multi-Peak Fat Water quantification. (a) IDEAL [13], (b) Yu et al. [17], (c) Berglund et al. [82], (d) Graph-Cut [15], (e) $B_0$-NICE [116], (f) GOOSE [117], (g) Hierarchical IDEAL [131], (h) 1D Mixed Fitting [16], (i) OSCAR approach. For each method, the cumulative MSE, $E_\xi$, is provided in brackets. White arrows in (a), (b), (d), (e) and (g) indicate the occurrence of significant errors (in terms of MSE). [Color figure can be viewed in the online issue].

**Fig. 4**. A comparison in terms of MSE for a 1.5 T MRI dataset of a FSHD subject under investigation among current reference implementations and our method. (a) IDEAL, (b) Yu et al., (c) Berglund et al., (d) Graph-Cut, (e) $B_0$-NICE, (f) GOOSE, (g) Hierarchical IDEAL, (h) 1D Mixed Fitting, (i) OSCAR approach. For each method, the cumulative MSE, $E_\xi$, is provided in brackets. White arrows in (a), (e) and (h) indicate the occurrence of significant errors (in terms of MSE). [Color figure can be viewed in the online issue].

**Fig. 5**. A comparison in terms of MSE for a 1.5 T MRI dataset from a NMD patient (having severe fat infiltrations in left thigh) among reference implementation and our proposed approach. (a) IDEAL, (b) Yu et al., (c) Berglund et al., (d) Graph-Cut, (e) $B_0$-NICE, (f) GOOSE, (g) Hierarchical IDEAL, (h) 1D Mixed Fitting, (i) OSCAR approach. For each method, the cumulative MSE, $E_\xi$, is provided in brackets. White arrows in (a), (e) and (h) indicate the occurrence of significant errors (in terms of MSE). [Color figure can be viewed in the online issue].

**Fig. 6**. A comparison in terms of MSE for a 1.5 T thigh MRI dataset from a NMD patient with severe disease conditions (complete atrophy and fatty transformation of the thigh muscles) among reference implementation and our proposed approach. (a) IDEAL, (b) Yu et al., (c) Berglund et al., (d) Graph-Cut, (e) $B_0$-NICE, (f) GOOSE, (g) Hierarchical IDEAL, (h) 1D Mixed Fitting, (i) OSCAR approach. For each method, the cumulative MSE, $E_\xi$, is provided in brackets. White arrows in (a), (c) and (h) indicate the occurrence of significant errors (in terms of MSE). [Color figure can be viewed in the online issue].

**Fig. 7.** Reproducibility study from six NMD subjects which were re-evaluated after a 1-week period. A $T_1$-weighted FSE MRI data (first column) is provided for reference purposes. The following muscular groups are considered: (Anterior muscular compartment) vastus lateralis (VL), vastus medialis (VM), (Posterior muscular compartment) Adductor (AD), Biceps Femoris (BF), Semimembranosus (SM), Semitendinosus (ST). For each candidate, a ROI is defined for above muscles with a different colored line (red line for VL, green line for VM, yellow line for AD, blue line for BF, cyan line for SM, purple line for ST) and average PDFF value indicated. The upper leg PDFF maps as obtained from the first and second scan (second and third column, respectively) were arranged and quantitatively compared in a range between 0 to 100%. [Color figure can be viewed in the online issue].

**Table 1**. Subjects Characteristics. Note: 'NA' means Not Applicable', 'Amb' means ambulant, 'wcd' means 'wheelchair dependent'. The following classification is applied to classify disease severity of pathological fatty infiltrations: Minimal (FF<=5%), Moderate (FF<=30%), Medium (30<FF<=50%), Severe (FF>50%). Note: A subset of candidates composed by Pompe (1) and FSHD (5) patients has been re-evaluated to analyze reproducibility performance.

**Table 2**. The total ICC of the upper leg muscles as shown in Fig. 7 for intra-scanner proton density fat-fraction (PDFF) imaging. The disease severity is reported in brackets under the description, by considering the degree of involvement of muscular compartments and following also the classification in Table 1.

**Table 3**. List of Different Multipeak CSE methods used in this study. The symbol "†(X)" indicates the number of fat peaks used by the model.

**Supplemental Materials - Caption Descriptions**

**Fig. 3.2** - A comparison in terms of PDFF for the public 3 T abdominal MRI dataset as already shown in Fig. 3, among current implementation of most of state-of-the-art algorithms for Chemical Shift Multi-Peak Fat Water quantification. (a) IDEAL, (b) Yu et al., (c) Berglund et al., (d) Graph-Cut, (e) $B_0$-NICE, (f) GOOSE, (g) Hierarchical IDEAL, (h) 1D Mixed Fitting, (i) OSCAR approach.

**Fig. 3.3** - A comparison of $B_0$ field inhomogeneity maps, using $f_B$ (in Hz), for the public 3 T abdominal MRI dataset as already shown in Fig. 3, among current implementation of most of state-of-the-art algorithms for Chemical Shift Multi-Peak Fat Water quantification. (a) IDEAL, (b) Yu et al., (c) Berglund et al., (d) Graph-Cut, (e) $B_0$-NICE, (f) GOOSE, (g) Hierarchical IDEAL, (h) 1D Mixed Fitting, (i) OSCAR approach.

**Fig.4.2** - A comparison in terms of PDFF for the 1.5 T thigh MRI dataset (same as in Fig. 4) among current reference implementations of most of state-of-the-art algorithms for Chemical Shift Multi-Peak Fat Water quantification. (a) IDEAL, (b) Yu et al., (c) Berglund et al., (d) Graph-Cut, (e) $B_0$-NICE, (f) GOOSE, (g) Hierarchical IDEAL, (h) 1D Mixed Fitting, (i) OSCAR approach.

**Fig.4.3** - A comparison in terms of $B_0$ field inhomogeneity maps, using $f_B$ (in Hz), for the 1.5 T thigh MRI dataset (same as in Fig. 4), among current implementation of most of state-of-the-art algorithms for Chemical Shift Multi-Peak Fat Water quantification. (a) IDEAL, (b) Yu et al., (c) Berglund et al., (d) Graph-Cut, (e) $B_0$-NICE, (f) GOOSE, (g) Hierarchical IDEAL, (h) 1D Mixed Fitting, (i) OSCAR approach.

**Fig.4.4** - Result panel for the 1.5 T thigh MRI dataset (same as in Fig. 4) with clusterization results. (a) Gap function $G_\Gamma$ for different number of clusters $i$. Sum of intra-cluster and inter-cluster distances as a function of $i$ from the same clustering method (left and right insets, respectively). (b) The magnitude for the first echo from the input data set. (c) The resulting clusterization into $\Gamma$ groups. (d) Spectral information as estimated using OSCAR (colored bars, one spectrum for each PA) and reference [17] pre-calibration approach (black bars). [Color figure can be viewed in the online issue].

**Fig.5.2** - A comparison in terms of PDFF for the 1.5 T thigh MRI dataset (same as in Fig. 5) among current reference implementations of most of state-of-the-art algorithms for Chemical Shift Multi-Peak Fat Water quantification. (a) IDEAL, (b) Yu et al., (c) Berglund et al., (d) Graph-Cut, (e) $B_0$-NICE, (f) GOOSE, (g) Hierarchical IDEAL, (h) 1D Mixed Fitting, (i) OSCAR approach.

**Fig.5.3** - A comparison in terms of $B_0$ field inhomogeneity maps, using $f_B$ (in Hz), for the 1.5 T thigh MRI dataset (same as in Fig. 5), among current implementation of most of state-of-the-art algorithms for Chemical Shift Multi-Peak Fat Water quantification. (a) IDEAL, (b) Yu et al., (c) Berglund et al., (d) Graph-Cut, (e) $B_0$-NICE, (f) GOOSE, (g) Hierarchical IDEAL, (h) 1D Mixed Fitting, (i) OSCAR approach.

**Fig.5.4** - Result panel for the 1.5 T thigh MRI dataset (same as in Fig. 5) with clusterization results. (a) Gap function $G_\Gamma$ for different number of clusters $i$. Sum of intra-cluster and inter-cluster distances as a function of $i$ from the same clustering method (left and right insets, respectively). (b) The magnitude for the first echo from the input data set. (c) The resulting clusterization into $\Gamma$ groups. (d) Spectral information as estimated using OSCAR (colored bars, one spectrum for each PA) and reference [17] pre-calibration approach (black bars). [Color figure can be viewed in the online issue].

**Fig.6.2** - A comparison in terms of PDFF for the 1.5 T thigh MRI dataset (same as in Fig. 6) among current reference implementations of most of state-of-the-art algorithms for Chemical Shift Multi-Peak Fat Water quantification. (a) IDEAL, (b) Yu et al., (c) Berglund et al., (d)

Graph-Cut, (e) $B_0$-NICE, (f) GOOSE, (g) Hierarchical IDEAL, (h) 1D Mixed Fitting, (i) OSCAR approach.

**Fig.6.3** - A comparison in terms of $B_0$ field inhomogeneity maps, using $f_B$ (in Hz), for the 1.5 T thigh MRI dataset (same as in Fig. 6), among current implementation of most of state-of-the-art algorithms for Chemical Shift Multi-Peak Fat Water quantification. (a) IDEAL, (b) Yu et al., (c) Berglund et al., (d) Graph-Cut, (e) $B_0$-NICE, (f) GOOSE, (g) Hierarchical IDEAL, (h) 1D Mixed Fitting, (i) OSCAR approach.

**Fig.6.4** - Result panel for the 1.5 T thigh MRI dataset (same as in Fig. 6) with clusterization results. (a) Gap function $G_\Gamma$ for different number of clusters $i$. Sum of intra-cluster and inter-cluster distances as a function of $i$ from the same clustering method (left and right insets, respectively). (b) The magnitude for the first echo from the input data set. (c) The resulting clusterization into $\Gamma$ groups. (d) Spectral information as estimated using OSCAR (colored bars, one spectrum for each PA) and reference [17] pre-calibration approach (black bars). [Color figure can be viewed in the online issue].